\documentclass[12pt]{article}
\usepackage{graphics}
\usepackage{graphicx}
\usepackage{amsmath,amsfonts,amssymb,amsthm}
\usepackage{multicol}
\usepackage{color}
\usepackage{makeidx}

\begin{document}

\begin{center}{\bf Boltzmann's Entropy and Large Deviation Lyapunov Functionals
  for Closed and Open Macroscopic Systems}\end{center}
\bigskip

\begin{center}
Joel L. Lebowitz\\
Department of Mathematics and Physics\\
Rutgers University\\
110 Frelinghuysen Road \\
Piscataway, NJ 08854 USA 
\end{center}

\noindent Abstract: I give a brief overview of the resolution of the apparent problem of
reconciling time symmetric microscopic dynamic with time asymmetric
equations describing the evolution of macroscopic variables.  I then
show how the large deviation function of the stationary state of the
microscopic system can be used as a Lyapunov function for the
macroscopic evolution equations.
\begin{center}------------------------------------------------------------------------------\end{center} 
\begin{center}{Gather ye rosebuds while ye may,\\*
Old time is still a-flying;\\*
And this same flower that smiles today,\\*
Tomorrow will be dying.}

\textit{Robert Herrick, 1591-1674}
\end{center}
\medskip

\begin{center}------------------------------------------------------------------------------\end{center} 

\noindent {\bf Keywords:  time's arrow, entropy, large deviations,
  lyapunov \\functions.   PACS: 75.10.-b}
\medskip


\section{Time's Arrow}
Every bit of macroscopic matter is composed of an enormous number of
atoms which act as
quasi-autonomous units.  Taking these atoms as classical
particles moving according to non-relativistic Hamiltonian equations the
complete microscopic (or micro)state of an isolated classical system of $N$
particles is specified by a point $X$ in its phase space
$\varGamma$,

\begin{equation}
X = ({\bf r_{1}}, {\bf v_{1}}, ... {\bf r}_{N}, {\bf v}_{N}),
\hspace{0.3cm} {\bf r}_{i} \in V \subset {\bf R}^{d}, \hspace{0.3cm}{\bf v}_{i} \in {\bf R}^{d}
\end{equation}
 and $X(t) = \phi_{t} X(0)$, where $\phi_{t}$ is the evolution operator
corresponding to the solution of the
Hamiltonian equations of motion.  These have the well known time
reversal symmetry:  setting, $RX = ({\bf r_{1}}, {\bf -v}_{1}, ..., {\bf r}_{N},
{\bf -v}_{N})$ then 
\begin{equation}
RX = \phi_{t} R \phi_{t} X, \hspace{0.3cm} t \in
(- \infty, \infty)
\end{equation}
Suppose now that some property of the system, specified by a function
$f(X(t))$, behaves in a particular way as $t$ increases, then there is also
a trajectory in which the property behaves in the time reversed way.  Thus,
for example, if particle densities get more uniform, say in a way described
by the diffusion equation, then there will also be evolutions in which the
density gets more nonuniform.  So why is one type of evolution,
corresponding to ``entropy'' increase in accord with the second ``law'',
common and the other never seen?

This problem was clearly stated by W. Thomson (later Lord
Kelvin) who wrote in 1874 \cite{T}:

``The essence of Joule's discovery is the subjection of physical
phenomena to dynamical law.  If, then, the motions of every particle
of matter in the universe were precisely reversed at any instant, the
course of nature would be simply reversed for ever after.  The
bursting bubble of foam at the foot of a waterfall would reunite and
descend into the water; the thermal motions would reconcentrate their
energy, and throw the mass up the fall in drops re-forming into a
close column of ascending water. ... And if also the materialistic
hypothesis of life were true, living creatures would grow backwards,
with conscious knowledge of the future, but no memory of the past, and
would become again unborn.  But the real phenomena of life infinitely
transcend human science; and speculation regarding consequences of
their imagined reversal is utterly unprofitable.''

The explanation of this apparent paradox, due to Thomson, Maxwell and
Boltzmann, is based on the great disparity between microscopic and
macroscopic scales---with the consequent exponentially large ratios 
between the number of microstates (phase space volume) corresponding to the 
different macrostates---and the fact that events are determined not
only by differential equations, but also by initial conditions.  These
provide the ingredients for the emergence of definite time asymmetric
behavior in the observed evolution of macroscopic systems, despite the total
absence of such asymmetry in the dynamics of individual atoms. (For a
more detailed discussion and references see [2],[3].)

\subsection{Macrostates}

To describe the macroscopic state of a system of $N$ atoms in a box
$V$, say $N \gtrsim10^{20}$, we make use of a much cruder description
than that provided by the microstate $X$.  We shall denote by $M(X)$ such
a macroscopic description of a macrostate. As an example we may divide 
$V$ into $K$ cells,
where $K$ is large but still $K \ll N$, and specify the number of
particles, the momentum and the amount of energy in each cell, with
some tolerance.

Clearly there are
many $X's$ (in fact a continuum) which correspond to the same $M$. Let
$\varGamma_{M}$ be the region in $\varGamma$ consisting of all microstates
$X$ corresponding to a given macrostate $M$ and denote by
$|\varGamma_{M}|$ its Liouville volume.

The equilibrium macrostate $M_{eq}$ is defined as that state for
which $|\varGamma_{Meq}| \sim |\varGamma_{E}|$, the area of the whole energy
surface.  When
$M(X)$ specifies a nonequilibrium state, $|\varGamma_{M(X)}|$ is much smaller.
Thus if the system contains $N$ atoms in a volume $V$ then the ratio of
$|\varGamma_{M_{eq}}|$, for the macrostate $M_{eq}$ in which there are 
$(\frac{1}{2} \pm 10^{-10})N$ particles in the left half of the box, to
$|\varGamma_M|$ for a macrostate $M$ in which all the particles are in the
left half is of order $2^N$.  For any macroscopic value of $N$, this is
far larger than the ratio of the volume of the known universe to the volume
of one proton. \footnote
{This is the reason why properties of an equilibrium system, such as the
fraction of particles in a given velocity domain 
can be obtained, for $N\gg1$, as an average over the
microcanonical ensemble.  N.B. This does not depend on
the system being ergodic in the mathematical sense as long as $N$ is
large enough.}

Boltzmann then argued that given this disparity in the sizes of $\varGamma_{M}$,
$|\varGamma_{M}(X(t))|$ will {\it typically} increase
in a way which  {\it explains} and describes the evolution  towards
equilibrium of isolated microscopic systems. 'Typical' here
means that for any $\varGamma_{M}$ the relative volume of the set
of microstates $X$ in $\varGamma_{M}$ for which $\log |\varGamma_{M}(X(t))|$ decreases by a
macroscopic amount during some time period $\tau$, (no bigger than the
age of the universe) goes to zero exponentially in the number of
atoms in the system.

\section{Entropy and Lyapunov Function for Isolated System}
To make a connection with the Second Law of Clausius,
Boltzmann defined the entropy of a macroscopic system with microstate
$X$ as 
\begin{equation}
S_B(X) = k \log|\varGamma_{M(X)}|
\end{equation}
and showed that (for a dilute gas) in the equilibrium macrostate
$M_{eq}$, i.e. $X \in \varGamma_{Meq}$,
$S_B$ is equal (to leading order in $N$) to the thermodynamic entropy of
Clausius. Following O. Penrose, I shall call $k{\rm log}$
$|\varGamma_{M}(X)|$ the Boltzmann entropy of a system in the macrostate $M(X)$.We can make Boltzmann's argument quantitative if we suppose (assume) that the time evolution of $M_t$ satisfies an autonomous
deterministic equation, such as the Navier-Stokes equation or the
Boltzmann equation. This means that if $M_{t_1} \rightarrow M_{t_2}$, 
then the microscopic dynamics $\phi_{t}$ carries
$\varGamma_{M_{t_1}}$ inside $\varGamma_{M_{t_2}}$,
i.e. $\phi_{t_2-t_1}\varGamma_{M_{t_1}}\subset\varGamma_{M_{t_2}}$ with
{\it negligible error}.  The fact that phase space volume is
conserved by the Hamiltonian time evolution implies that
$|\varGamma_{M_{t_1}}|\leq|\varGamma_{M_{t_2}}|$ and thus that
$S_B(M_{t_2})\geq S_B(M_{t_1})$ for $t_2\geq t_1$.

We have thus derived an ``$\mathcal{H}$-theorem'' or Lyapunov function for any
deterministic evolution of the macro-variables arising from the
microscopic dynamics of an isolated Hamiltonian system[3].

{\bf Example}:  For spatially uniform equilibrium systems the
  thermodynamic entropy is extensive
\begin{equation}
S(E, {\bf N}, V) =  V s(e,{\bf n}).
\end{equation}

$s(e,.)$ is a concave function of $e$.

\begin{equation}
\frac{\partial s}{\partial e} = \frac{1}{T},\hspace{.5cm}
\frac{\partial}{\partial e}(\frac{1}{T}) = -(1/T^{2})\frac{\partial T}{\partial e}\leq 0.
\end{equation}

For systems in ``local thermal equilibrium'' (LTE)
with local densities $n(x), e(x), {\bf u(x)}$

\begin{equation}
S_{B}(n,{\bf u},e) = \int_V s(e({\bf x}) - \frac{1}{2} mn({\bf x})
{\bf u}^2 ({\bf x}), n({\bf x}))d{\bf x} = S_{l.e}
\end{equation}

Consider now an 
isolated system in LTE (with ${\bf u}=0$ and $n$ constant) in a region
$V$ with boundary surface $\sum$ and an
energy density profile $e({\bf x})$ satisfying the macroscopic conservation
equation

\begin{equation}
\frac{\partial e}{\partial t} = -\nabla \cdot {\underbar{J}}
\end{equation}

where $\bf J(e)$ is the heat flux.  When this is given by Fourier's law,  

\begin{equation}
\bf J = -\kappa \nabla T, \quad \kappa(T) \ge 0.
\end{equation}

we then have a closed autonomous equation for $e$ or $T$.  This yields,

\begin{equation}
\begin{gathered}
\frac{dS_{l.e}}{dt} = \frac{d}{dt} \int_V s d{\bf x}\\ = -
\int_{V}\frac{1}{T} (\nabla \cdot {\bf J}) d x
 = - \int_{\sum}\frac{1}{T} \bf J \cdot d{\sum} + \int_{V} \bf J
\cdot (\nabla\frac{1}{T}) {\bf d x} \ge 0,
\end{gathered}
\end{equation}

\noindent since $\bf J \cdot {d\sum} = 0$ and $\kappa \geq 0$. 

We next consider what happens when the isolated system is not in local
equilibrium. (Following that we shall consider situations when the
system is not isolated.)

Following Boltzmann, we refine the thermodynamic $M$ used for systems
in LTE 
 by noting
that the microstate $X =
\{{\bf r}_i, {\bf v}_i\}$, $i = 1,...,N$, can be considered as a set of
$N$ points in the six dimensional ``$\mu$-space''.  We then divide up 
this $\mu$-space into $\tilde J$ cells
$\tilde \Delta_\alpha$, centered on $({\bf r}_\alpha, {\bf v}_\alpha)$, of
volume $|\tilde \Delta_\alpha|$.  A macrostate $\tilde M$ is then
specified by the number of particles in each $\tilde\Delta_\alpha$, 
\begin{equation}
\tilde M = \{N_\alpha\}, \quad \alpha = 1,...,\tilde J << N.
\end{equation}

For dilute gases one can {\it neglect}, for typical configurations,
the interaction energy between the particles.
The coarse grained energy of the system in the state
$\tilde M$ is given, up to terms independent of $f$, by
\begin{equation}
\frac{1}{2} m \sum_\alpha N_\alpha {\bf
v}^2_\alpha = E 
\end{equation}
with 
\begin{equation}
\sum N_\alpha = N
\end{equation}

The phase space volume associated with such an $\tilde M$ 
is then readily computed to be 
\begin{equation}
|\varGamma_{\tilde M}| = \Pi_\alpha(N_\alpha!)^{-1}
|\tilde \Delta_\alpha|^{N_\alpha} 
\end{equation}

Stirling's formula then gives
\begin{equation}
S_B(\tilde M) \sim  -k \bigg\{\sum_\alpha \Bigl(\frac{N_\alpha}{|\tilde \Delta_{\alpha|}} \log \frac{N_\alpha}{|\tilde \Delta_{\alpha|}}\Bigr)
|\tilde \Delta_\alpha|- N \bigg \}
\end{equation}
Using $\tilde M$ we can associate with a typical $X$ a coarse grained
density\\ $f_X \sim N_\alpha/|\tilde \Delta_\alpha|$ in $\mu$-space,
i.e.\ such that $N_\alpha =
\int_{\tilde \Delta_\alpha} d{\bf x} d{\bf v} f_X({\bf x},{\bf v})$. 
The Boltzmann entropy is then given by 
\begin{equation}
S_B(X) = S_{\rm gas} (f) = -k \int_V d{\bf x} \int_{{\mathbb{R}}^3} d{\bf
v} f({\bf x},{\bf v}) \log f({\bf x},{\bf v})   
\end{equation}

The 
maximum of $S_{\rm gas}(f)$ over all $f$ which satisfy the constraints,
\begin{equation}
\int_V d{\bf x} \int_{{\bf R}^3} d{\bf v} f({\bf x}, {\bf v}) = N 
\end{equation}
\begin{equation}
\int_V d{\bf x} \int_{{\Bbb R}^3} d{\bf v} \,\frac{1}{2} m{\bf v}^2
f({\bf x}, {\bf v}) = E 
\end{equation}
gives the equilibrium distribution, which is readily seen to be the
Maxwell distribution
\begin{equation}
f_{eq} = \frac{N}{V}(2\pi k T/m)^{-3/2} \exp[- m{\bf v}^2/2kT] 
\end{equation}

\noindent where $kT = 2/3 (E/N)$.  In this case $S_{B}$ coincides with the Clausius entropy\endgraf

\begin{equation}
S_{\rm gas}(f_{eq})=S(E,N,V)=Nk[\frac{3}{2}\log T-\log(N/V)] 
+ {\rm Const.}
\end{equation}

When $f \ne f_{eq}$ then $f$ and consequently $S_{\rm gas}(f)$ will
change in time.  The second law, now says
that {\it typical $X \in \varGamma_{\tilde M}$}, at the initial time $t=0$, will
have an $\tilde M_t = \tilde M(X_t)$ such that
$S_B(\tilde M(X_t)) \geq S_B(\tilde M(X_{t^\prime}))$, for $t \geq
t^\prime$.  This means that $S_{\rm gas}(f_t) \geq S_{\rm
gas}(f_{t^\prime})$, for $t \geq t^\prime$.  This is exactly what
happens for a dilute gas described by the Boltzmann equation for which 

\begin{equation}
\frac{d}{dt} S_{\rm gas} (f_t) \geq 0, \hspace{0.3cm} {\rm
  Boltzmann's}\hspace{0.3cm} {\cal H}{\rm -theorem}
\end{equation}
i.e. $S_{\rm gas}(f)$ is a Lyapunov  function.

As put by Boltzmann:
``In one respect we have even generalized the entropy principle here,
in that we have been able to define the entropy in a gas that is not
in a stationary state''[4].

{\bf Remark}: It is important to distinguish between the empirical $\mu$-space
density profile $f_{X_t}({\bf x}, {\bf v})$ and
another object with the same name, the marginal one-particle
(probability) distribution $F_1({\bf x}, {\bf v}, t)$ obtained from an
$N$-particle ensemble density evolving according to the Liouville
equation. An instructive example is 
a macroscopic system of $N$ noninteracting point particles,
moving among a periodic array of scatterers in a macroscopic volume
$V$.  Starting with a nonuniform initial density
$f_{X_0}({\bf x}, {\bf v})$ the time evolved $f_{X_t}({\bf x}, {\bf
v})$ will approach an $f$ which depends only on $|{\bf v}|$ and which
will have a larger $S_{\rm gas}(f)$, while
$\int\int F_1 \log F_1 d{\bf x} d{\bf v}$ remains constant in time.
The 
obvious evolution equation for $f_{X_t}$ for this system, namely the
one-particle Liouville equation, in fact does not describe the
evolution of $f_{X_t}$ for times after which $F_1({\bf x}, {\bf v},
t)$ has developed structure on the microscopic scale.

\subsection{The Boltzmann Entropy of Dense Fluids Not in LTE}

Consider now the case
when the interaction potential energy $\Phi$ between the particles is not
negligible.  The region $\varGamma_{\tilde M}$ will then include
phase points with widely differing total energies.  The set of
microstate $X$ of a system with a specified energy, $H(X) = E$ will
then correspond to a small fraction of $\varGamma_{{\tilde M}(X)}$.
In fact a little thought shows that most of $\varGamma_f$ corresponds to
the largest energies compatible with $f({\bf x},{\bf v})$. The
macrostate $M$ specified by  both $f$ and $E$ will then have a
Boltzmann entropy consisting of a momentum part and a configurational part.
For a system of hard spheres where $E=K$ the
Boltzmann entropy can be written as the sum,
\begin{equation}
S_{hs}(f) = S^{(m)}(f) + {\cal S}^{(c)}_{hs}(n)
\end{equation}
where $S^{(m)}$ is the momentum part 
\begin{equation}
S^{(m)}(f) = -\int\limits_{V}d{\bf x}\int d{\bf v} f ({\bf x,v})
\log[f({\bf x,v})/n({\bf x})]
\end{equation}
and $\mathcal{S}^{(c)}_{hs}(n)$ is the
configurational part of the entropy of an equilibrium system of hard
spheres kept at a nonuniform density $n({\bf x}) = \int 
f({\bf x},{\bf v}) d{\bf v}$ by some external potential $U({\bf x})$.

$S_{hs}(f)$ was proven by
Resibois (in a different form) to be a Lyapunov function for
the modified Enskog equation c.f.[3], 

\begin{equation}
\frac{d}{dt} S_{hs}(f_t) \geq 0  
\end{equation}

\includegraphics[width=1\textwidth]{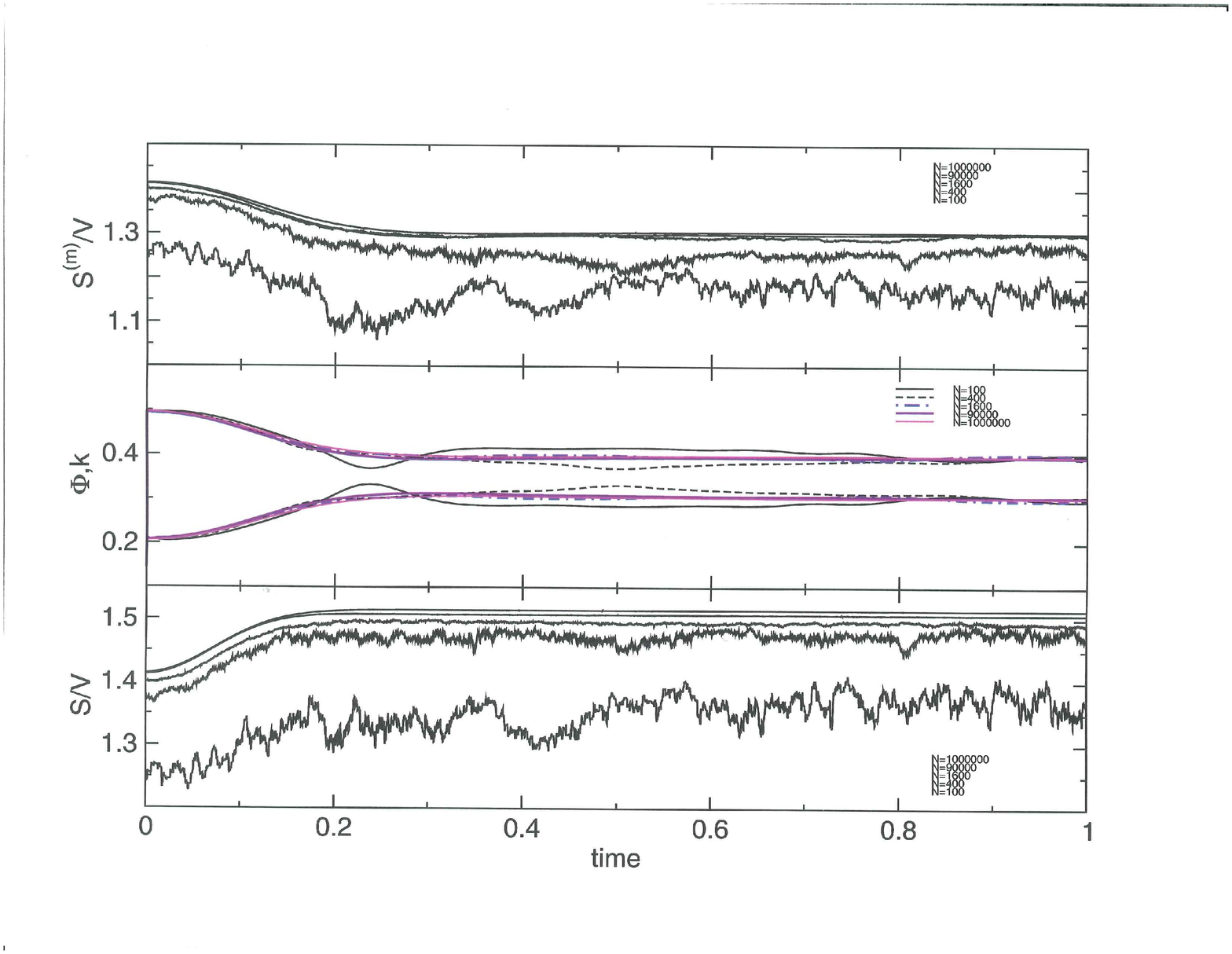}

The decrease of $S^{(m)}(f_{t})$ and increase of the total $S(f_{t}, E)$
is shown in Fig. 1.  Based on molecular dynamic simulations for
different number of particles $N$ in a periodic box, Fig.1 shows the time evolution of $S^{(m)}$, of the potential and
kinetic energies and of the total entropy $S$ for a Lennard-Jones
system started at $t = 0$ in a state where the kinetic energy is ``too
high''[5].  This corresponds to a situation considered by Jaynes, c.f.[3].

\section {Open Systems}
For an open system, say one in contact with a ``heat bath'' at a
specified temperature $\bar T$, the entropy of  the system
alone is clearly no longer an increasing function, e.g. we can start
the system at an energy $E_{o}$ corresponding to a temperature $T_{o}
> \bar T$. In terms of the macroscopic equation for the energy density
or temperature the entropy is no
longer a Lyapunov function since the energy flux across the boundaries
no longer vanishes, and can be either positive or negative.

So what do we do for a Lyapunov function?

A simple way which works for the case when there is only one heat bath
is to observe that the
{\it total} entropy production in system plus reservoir can be
written as, see (9),

\begin{equation}
\frac{d S_{total}}{d l} = \frac{d S_{l.e}}{d t} + \int_{\sum}
(1/\bar{T}) \bf J \cdot {d \sum}
\end{equation}

\begin{equation}
= \frac{d}{dt}[S_{l.e} - (1/\bar T) E_{l.e}] = \frac{d}{dt}(\mathcal{-F})
\end{equation}

\begin{equation}
= \int_{V} \bf J \cdot \nabla(\frac{1}{T})dx \ge 0
\end{equation}
where $\mathcal{F}$, given by the terms in the square brackett in (25), is now the ``Lyapunov function''.

This procedure fails when the system is in contact with more than one
heat bath and $\bar T$ is not constant on the boundary $\sum$ in which
case the entropy production is not zero in the stationary state.

To proceed we now recall that, as noted by Boltzmann and Einstein, the
relative Boltzmann entropy $\mathcal{S}(M)\equiv
S_{B}(M)-S_{B}(M_{eq})$ is equal to the log of the probability of
finding the system in the macrostate $M = \{e(x)\}$,

\begin{equation}
P(M) \sim \exp [S_{B}(M) - S_{B}(M_{eq})] = \exp \{-\mathcal{F}(M)\}
\end{equation}

This probability is with respect
to the uniform (microcanonical) measure on the energy surface of the isolated system,
which is stationary under the
microscopic Hamiltonian time evolution.

$\mathcal{S}(M)$ thus coincides, in the limit of large system size and
$M$ macroscopically distinct from $M_{eq}$ (the latter includes states
which only differ by ``normal'' fluctuations) with the negative of the
usual large deviation functional (LDF) of probability theory for
$\mu_{st}(X)\sim \delta (H(X)-E)$, i.e. for the microcanonical ensemble.

The same is true for the Lyapunov function $\mathcal{F}$ for the system in contact with
a single heat bath at temperature $\bar T$, where
\begin{equation}
\mathcal{F}(\{e(x)\}) = [E - \bar{T} S_{l.e}]/ \bar{T}
\end{equation}
is again the LDF of the stationary measure for the system in contact
with a heat bath.  This is now the canonical ensemble at temperature
$\bar T$.

\begin{equation}
\mu_{st} \sim exp [- H (X)/ \bar T]
\end{equation}

The above analysis can be readily generalized to the
macrostate $M = \{e(\bf x), n(\bf x), \underline{u}(\bf x)\}$ whose time evolution is
governed by the Navier-Stokes equations.  In fact one expects that the
LDF for the stationary measure will always be a Lyapunov function for
the macroscopic equation [6, 7].  An example which exploits this fact to
derive new Lyapunov functions is given in the next section[8].

\section {Lyapunov function for a system in contact with several particle reservoirs}

Let $\sigma$ be a smooth increasing function.
We consider the PDE on a regular domain $V \subset \mathbb{R}^d$
\begin{equation}
\frac{\partial \rho(t,\bf x)}{\partial t}= \nabla^{2}\Big( \sigma \big( \rho(t,x) \big) \Big) \, ,
\end{equation}
with Dirichlet boundary conditions on $\Sigma$ specified by
the reservoirs, i.e. $\rho(t,x) = \bar{\rho}(x)$ for $x \in \Sigma$ where $\bar{\rho}(x)$ is the stationary profile in all of $V$.
Let
\begin{equation}
F_u (v) = \int_u^v dz \, \log \frac{\sigma(z)}{\sigma(u)} \, .
\end{equation}
We define the functional

\begin{equation}
\mathcal{F} (\sigma) = \int_{V} dx \, F_{\bf u)} (\sigma(\bf x)).
\end{equation}

A straightforward but lengthy computation then shows that
$\frac{\partial \mathcal{F}(\rho)}{\partial t} \ge 0$ where we used the fact that
on the boundary  $\frac{\rho(t,x)}{\bar{\rho}(x)} = 1$.\\

The function $\mathcal{F}$ defined in (32) is the LDF for the ``zero range process'' in
contact with particle reservoirs at different densities.   For this
model the NESS is known explicitely and thus permits the explicit
computation of the LDF [8].

The same computation will go through with a field

\begin{equation}
\frac{\partial_{\rho}(t,x)}{\partial_{t}} = \nabla^{2} \sigma (\rho(t,{\bf x})) - E  \nabla \sigma (\rho(t,\bf x)).
\end{equation}

One could also treat mixed Dirichlet/Neumann boundary conditions.

I thank the organizers of this conference for a wonderful meeting.
Much of the work desribed here was done jointly with S. Goldstein.
Work supported by NSF grant DMR08-02120 and AFOSR grant AF-FA 09550-07.

\end{document}